# NGC 4526 gas, High Velocity Clouds, and Galactic Halo gas: the Interstellar Medium towards SN 1994D[*]

D. L. King[1], G. Vladilo[2], K. Lipman[3], K. S. de Boer[4], M. Centurión[5], P. Moritz[6], N. A. Walton[7]

[1] Royal Greenwich Observatory, Madingley Road, Cambridge, CB3 0EZ, United Kingdom
[2] Osservatorio Astronomico di Trieste, Via G.B. Tiepolo 11, 34131 Trieste, Italy
[3] Institute of Astronomy, Madingley Road, Cambridge, CB3 0HA, United Kingdom
[4] Sternwarte, University of Bonn, Auf dem Hügel 71, D-53121, Bonn, Germany
[5] Instituto de Astrofísica de Canarias, Via Lactea s/n, 38200 La Laguna, Tenerife, Spain
[6] Radioastronomisches Institut, University of Bonn, Auf dem Hügel 71, D-53121, Bonn, Germany
[7] Royal Greenwich Observatory, Apartado 321, Santa Cruz de La Palma, 38780 Tenerife, Spain

**Abstract.** We present spectroscopic observations of supernova 1994D in NGC 4526, an S0$_3$ galaxy in the Virgo cluster 15 Mpc distant. The datasets consist of the interstellar Ca II and Na I lines towards the supernova at high spectral resolution (FWHM 6 km s$^{-1}$), H $\alpha$ and [N II] observations at lower resolution (FWHM 33 km s$^{-1}$) of the nucleus of NGC 4526 and the supernova, obtained with the William Herschel Telescope at La Palma, and 21 cm spectra obtained with the 100 m Effelsberg Radiotelescope in the field of NGC 4526.

The velocity of the gas in NGC 4526 determined from our H $\alpha$ spectra is +625 km s$^{-1}$ at the centre (systemic velocity) and +880 km s$^{-1}$ at the supernova position. Our value of the systemic velocity is higher than the value of +450 km s$^{-1}$ frequently quoted in the literature.

In our high resolution spectra we detect Ca II and Na I absorption at +714 km s$^{-1}$ which is produced in interstellar gas in NGC 4526. To our knowledge this is the first detection of interstellar absorption originating in a galaxy of early morphological type. The ratio $N(\mathrm{Na}^0)/N(\mathrm{Ca}^+) \simeq 4$ suggests an origin in cold gas at rest velocity relative to its galactic environment. The lack of multiple components indicates a relatively simple structure of the interstellar medium in the inner region of NGC 4526, at least in the particular line of sight to the supernova at the border of the nuclear ring of dust.

We detect multi-component Ca II and Na I absorption lines in the range from +204 to +254 km s$^{-1}$ which originate in a complex of High Velocity Clouds (HVCs) located at a distance $\ll$ 1 Mpc, in the surroundings of the Milky Way. This rare detection of HVCs in absorption enables the study of the properties of the gas using the Ca$^+$ and Na$^0$ column densities, combined with the H$^0$ column density taken from the literature at +215 km s$^{-1}$ in the same line of sight. We find $N(\mathrm{Na}^0)/N(\mathrm{Ca}^+) = 0.1 - 0.4$, in our Galaxy the signature of high velocity gas. The Ca$^+$/H$^0$ and Na$^0$/H$^0$ column density ratios are extremely high compared to Milky Way interstellar values; the gas appears to have near solar abundances, very low dust content, and a diluted ultraviolet radiation field. This is entirely consistent with Galactic fountain models, in which hot gas is expelled into the outer halo, and subsequently cools.

At $-29$ km s$^{-1}$, we find weak Ca II absorption and weak H I emission. This component has properties similar to those of the warm gas around the Sun and may originate in gas infalling onto the Galactic disk, perhaps associated with the extended complexes of Galactic halo gas at intermediate negative velocities which are present in the northern Galactic hemisphere.

Finally, close to rest velocity, we find both warm and cold gas located beyond 65 pc, probably associated with high latitude gas at the border of Loop I.

The total reddening of the supernova, estimated using the standard Milky Way gas-to-dust ratio, is $E(\mathrm{B-V}) \simeq 0.05$.

Supernovae have been studied in depth due to their impact on many astrophysical problems, particularly enrichment of the interstellar medium, neutrino physics and heavy element synthesis. The advent of high resolution spectrographs and efficient detectors has, over the last decade, allowed us to study the interstellar and intergalactic gas in the direction of bright supernovae in external galaxies through absorption line spectroscopy with sensitivities previously unattainable. SN 1987A enabled the study of the interstellar medium in the direction of and within the Large Magellanic Cloud to a distance of 0.05Mpc (Vidal-Madjar et al. 1987, de Boer et al. 1987). SN 1993J extended the search for interstellar and intergalactic absorptions out to the distance of the M 81 group at 3.6 Mpc, where evidence of Galactic gas, intermediate velocity clouds, gas within the host galaxy and tidally stripped material in the intergalactic space within the M 81 group was found (Vladilo et al. 1993,1994, de Boer et al. 1993, Bowen et al. 1994). SN 1994D now allows us to probe gas out to much larger distances. SN 1994D ($l = 290.2°$, $b = 70.1°$) in NGC 4526, an $S0_3$ lenticular galaxy (Sandage 1961) and a member of the Virgo cluster of galaxies at a distance of 15 Mpc (Pierce et al. 1994) was discovered on March 7, 1994 (Treffers et al. 1994) using the automated Leuscher Observatory Supernova Search, about one week before maximum brightness. Until comparatively recently, early type galaxies were considered to be almost free of dust and gas. $S0_3$ galaxies by definition contain dust; some are now known to contain neutral hydrogen and molecular gas and have been detected at X-ray wavelengths (Roberts et al. 1991). The line of sight to SN 1994D is only 5° away from that towards SN 1991T in NGC 4527 which has been studied using interstellar Ca II absorption lines (Meyer & Roth 1991). We therefore have the opportunity to study gas along a similar sight line.

We present an analysis of high resolution observations of interstellar Ca II and Na I absorption lines towards SN 1994D. In addition, we present lower resolution H $\alpha$ and [N II] observations of NGC 4526 and 21 cm emission line profiles obtained in the field of NGC 4526. In the next section we describe the observations and line fitting analysis. We discuss the results in section 3.

## 2. Observations and data reduction

### 2.1. Optical spectra

The optical data were obtained using the William Herschel Telescope (WHT) at the Observatorio del Roque de Los Muchachos, La Palma, Spain.

The high resolution data were obtained with the Utrecht Echelle Spectrograph (Walker & Diego 1985) at the Nasmyth focus of the WHT. The detector was a cooled Tektronix CCD with 1024 x 1024, 24 micron square pixels. A spectrograph slit width of one arcsecond was used, $D_2$, $D_1$. Details of the observations are given in Table 1. At the time of the observations the visual magnitude was 12.5 (Lewis 1994). We also obtained Na I spectra of two bright foreground stars in the field of NGC 4526 to study the distribution of local gas close to the line of sight to SN 1994D. These stars are bright ($m_v \simeq 5$), with high rotational velocity; the spectra of these stars were used to correct for telluric absorption in the supernova Na I spectra.

In addition to the high resolution UES observations of SN 1994D, lower resolution spectra covering the H $\alpha$ and [N II] lines were obtained using the ISIS spectrograph (Carter et al. 1993), a 1200 lines per mm grating and an EEV CCD with 1242 x 1152, 22.5 micron square pixels, at the Cassegrain focus of the WHT. The one arcsecond wide slit, yielding a resolution of 33 km s$^{-1}$ at H $\alpha$, was placed across the nucleus of NGC 4526 and SN 1994D. Details of the observations are again given in Table 1.

Calibration frames – bias frames, flatfields at each instrumental setup and frequent arc exposures (thorium-argon hollow cathode lamp for the high resolution data, copper-neon hollow cathode lamp for the lower resolution data) to wavelength calibrate the data were also taken.

Table 1. Journal of the optical observations

| Date | Instrument | Target | Central $\lambda$ (Å) | Exposure (secs) |
|---|---|---|---|---|
| 09/3/94 | ISIS | SN 1994D | 6563 | 1750 |
| 15/3/94 | UES | " | 5900 | 10,800 |
| " | " | HD 110411 | " | 180 |
| " | " | HD 110423 | " | 360 |
| " | " | SN 1994D | 3955 | 1800 |
| " | " | " | 3953 | 1800 |
| 16/3/94 | " | " | " | 7200 |
| " | " | " | 3961 | 1800 |

All optical data sets were reduced using the Starlink FIGARO package (Shortridge 1990). The usual steps of bias subtraction, flatfielding, cosmic ray removal, wavelength calibration and sky subtraction were performed.

The hollow cathode lamp frames were reduced in an identical manner to the object frames. The spectral resolution determined from the profile of the arc lines was for both the UES and the ISIS data in good agreement with the nominal resolving power. The data were binned to two bins per resolution element.

The UES data were corrected for atmospheric absorption in the Na I region using the template spectrum from the foreground stars. The individual spectra were then corrected to a common LSR velocity scale; each spectral region was then co-added and normalised.

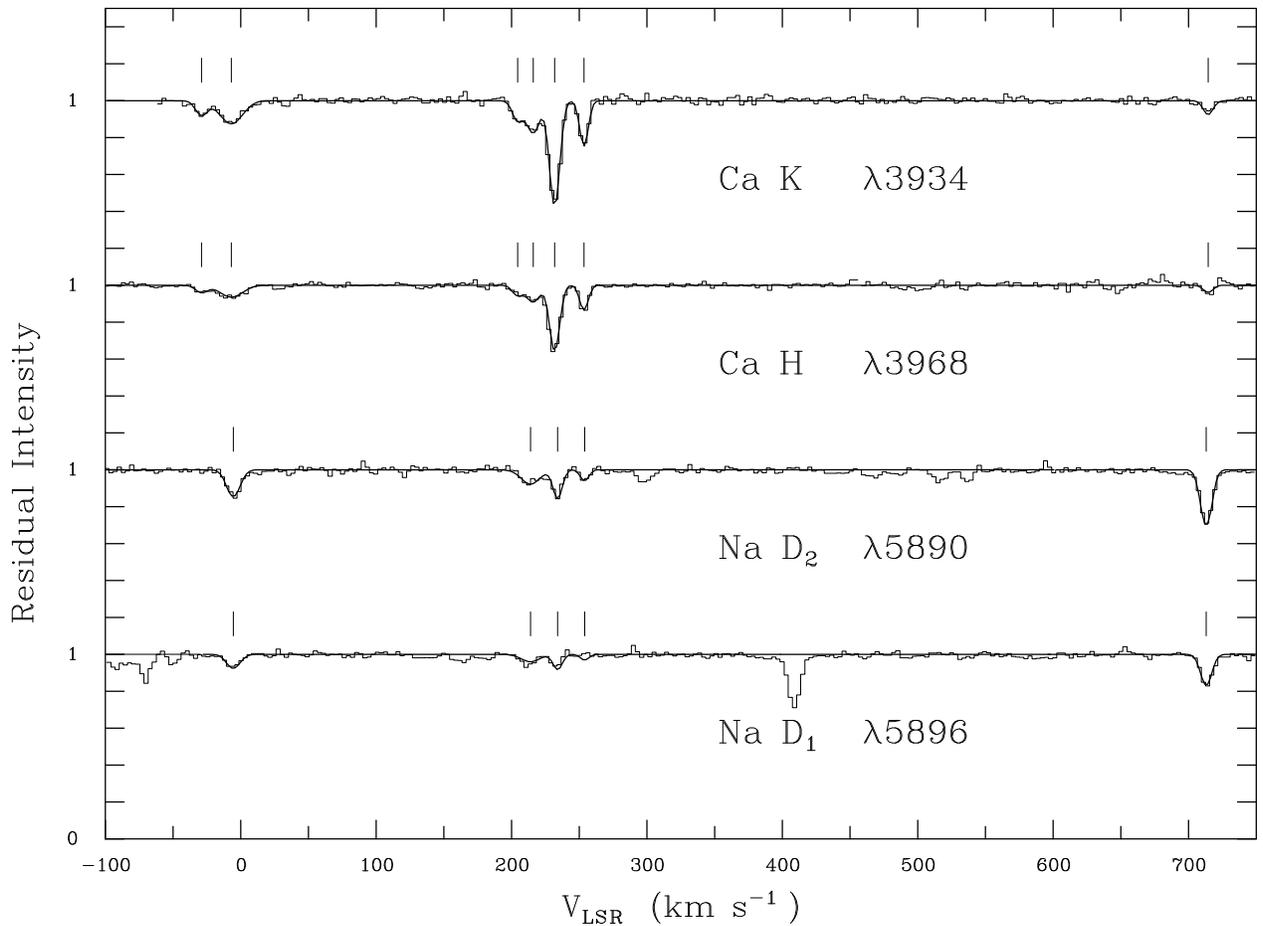

**Fig. 1.** Interstellar lines towards SN 1994D. Components from each doublet are indicated by tickmarks. Note that there is an overlap in wavelength between the Na I $D_2$, $D_1$ components. The observed data are plotted as a fine line, the synthetic profiles as a heavy line.

In Table 2 we present the signal-to-noise ratio in the spectral regions adjacent to the absorption features. We also give the minimum detectable equivalent width, calculated as $W_{min} = 3\Delta\lambda_{inst}/\frac{S}{N}$. These limits are converted to column density upper limits using the linear part of the curve of growth (Spitzer 1978).

**Table 2.** Data Characteristics

| Transitions | S/N | EW limit ($3\sigma$) (mÅ) | $N(X)$ limit ($3\sigma$) ($10^{10}$ cm$^{-2}$) |
|---|---|---|---|
| Ca II | 54 − 102 | 4.5 − 2.5 | 5 − 3 |
| Na I | 75 | 5 | 2.5 |

Note:− The Ca II H, K data were obtained with three different central wavelength settings leading to varying signal-to-noise ratios for these lines.

For the ISIS data, the seeing, determined from the profile of the guide star at the time of the observations, was 1.1 arcseconds; one dimensional spectra were created from the two dimensional sky subtracted spectrum by extracting in 4 spatial pixel increments (1 pixel = 0.33 arcseconds), centred on the nucleus of NGC 4526 and extending out to the supernova position.

### 2.2. Radio observations

The Effelsberg 100m Radiotelescope was used on April 8, 1994 to obtain 21 cm H I emission observations in the supernova field, one set centred on NGC 4526 and the second centred ten arcminutes due north. Spatial resolution (half power beam width ) at 21 cm is 9.1 arcminutes and the main beam efficiency $\eta_{mb} = 0.72$. The observations were performed using a cooled, dual channel FET amplifier (system temperature $T_{sys} \simeq 30$ K) and 1024 channel autocorrelator, split into two receivers with 512 channels of 1.29 km s$^{-1}$, and a velocity range of 660 km s$^{-1}$. Data were obtained at both positions centred on 450 km s$^{-1}$, with integration times of 1440 seconds, and at the first position centred on −50 km s$^{-1}$, with an integration time of 480 seconds. The frequency switching technique was used with reference frequency offsets by −3.5 MHz and +3.5 MHz for the velocity centres 450 km s$^{-1}$ and −50 km s$^{-1}$ respectively. All measurements were calibrated using the brightness temperature scale calibration point S7 (Kalberla et al. 1982) and corrected for stray radiation (Kalberla et al. 1980).

## 3. Analysis of the spectra

### 3.1. UES spectra

The Ca II and Na I interstellar spectra are shown in Fig. 1. Clearly, lines are blended; therefore profile fitting was performed by means of multi-component synthetic spectra. The analysis was performed using XVOIGT (Mar and Bailey 1994), where synthetic absorption profiles are compared with the observed profiles. Each component is reproduced with a Voigt profile, using the formalism described originally by Stromgren (1948) – in practice, three parameters describe the profile: radial velocity, broadening parameter, and column density. Combining these parameters with rest wavelength and oscillator strength (Morton 1991) enables the formation of the intrinsic absorption profile which, when convolved with the instrumental profile (determined from the FWHM of unblended lines in the thorium-argon spectrum), can be adjusted by iteration to obtain a best fit to the observed profile. A further constraint is achieved by fitting both components of each doublet simultaneously to derive self-consistent results. Results of the absorption line analysis are shown in Table 3.

A total of seven absorption components in Ca II and five in Na I are seen, covering the radial velocity range −29 km s$^{-1}$ ≤ $v_{lsr}$ ≤ +715 km s$^{-1}$ with good agreement between the two species. These break down into three distinct groups: those with velocities in the ranges −29 km s$^{-1}$ ≤ $v_{lsr}$ ≤ −5 km s$^{-1}$, +204 km s$^{-1}$ ≤ $v_{lsr}$ ≤ +254 km s$^{-1}$, and $v_{lsr}$ ≃ +714 km s$^{-1}$.

### 3.2. ISIS spectra

The H $\alpha$ and [N II] spectra, from the nucleus out to the supernova position, are shown in Fig. 2. Clearly, there is an increase in $v_{lsr}$ from the nucleus outwards, with a corresponding change in intensity. Fits to the emission line profiles were performed using the ELF routines contained in the Starlink DIPSO package (Howarth et al. 1994). The fit is performed by specifying an estimate of the central wavelength and width of each emission line; gaussians are then fitted to the observed profile and then optimised by spectrum in an iterative manner.

The systemic velocity of NGC 4526 as given by the lines formed in the nucleus extraction is +625 km s$^{-1}$. This increases to +880 km s$^{-1}$ at the supernova position. Note that close to the nucleus of NGC 4526 the lines are asymmetric.

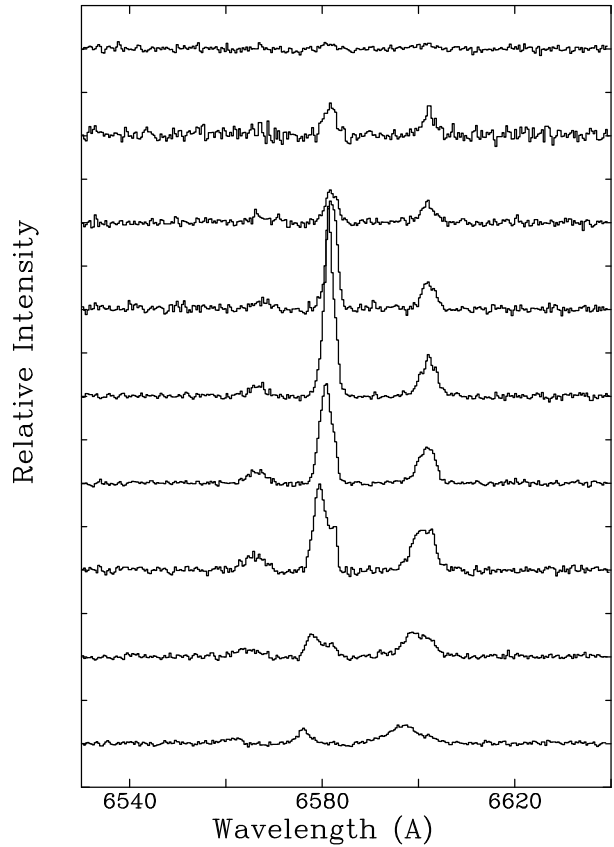

**Fig. 2.** H $\alpha$ and [N II] lines in NGC 4526. The spectra start at the nucleus of NGC 4526 at the bottom, incrementing in 1.3 arcsecond steps out to the supernova position in the top spectrum. The rest wavelength of H $\alpha$ is 6562.8 A and of [N II] are 6548.1 and 6583.4 Angstroms.

### 3.3. Effelsberg spectra

The 21 cm spectra of NGC 4526 are shown in Fig. 3. Gaussians were fitted to the emission line profiles using the CLASS package (Forveille et al. 1992). To be fitted adequately, the data required three components, the results of which are presented in Table 4. These components are in good agreement with the low velocity components seen

**Table 3.** Absorption Components towards SN 1994D

| Ca II H & K | | | Na I D$_2$,D$_1$ | | | |
|---|---|---|---|---|---|---|
| $v_{lsr}$ (km s$^{-1}$) | $b$ (km s$^{-1}$) | log N (cm$^{-2}$) | $v_{lsr}$ (km s$^{-1}$) | $b$ (km s$^{-1}$) | log N (cm$^{-2}$) | $\frac{N(Na^0)}{N(Ca^+)}$ |
| $-29.0^{+1.0}_{-1.0}$ | $5.0^{+1.0}_{-2.0}$ | $11.13^{+0.05}_{-0.11}$ | | | $< 10.40$ [1] | $< 0.19$ |
| $-7.0^{+1.0}_{-1.0}$ | $11.0^{+1.5}_{-1.5}$ | $11.62^{+0.03}_{-0.07}$ | $-5.5^{+1.0}_{-1.0}$ | $6.0^{+1.0}_{-1.0}$ | $11.30^{+0.03}_{-0.05}$ | 0.48 |
| $204.5^{+1.0}_{-1.0}$ | $4.0^{+2.0}_{-4.0}$* | $11.22^{+0.08}_{-0.05}$ | | | $< 10.40$ [1] | $< 0.15$ |
| $216.0^{+1.0}_{-1.0}$ | $5.5^{+1.5}_{-3.0}$ | $11.53^{+0.03}_{-0.05}$ | $214.0^{+2.0}_{-2.0}$ | $9.0^{+3.0}_{-5.0}$ | $11.15^{+0.05}_{-0.05}$ | 0.42 |
| $231.8^{+0.3}_{-0.3}$ | $3.5^{+0.5}_{-0.5}$ | $12.10^{+0.03}_{-0.05}$ | $234.0^{+1.0}_{-1.0}$ | $3.0^{+1.0}_{-1.0}$ | $11.17^{+0.01}_{-0.05}$ | 0.12 |
| $253.5^{+0.5}_{-0.5}$ | $2.5^{+1.5}_{-1.5}$ | $11.55^{+0.05}_{-0.03}$ | $254.0^{+1.0}_{-1.0}$ | $2.5^{+1.0}_{-2.0}$ | $10.70^{+0.07}_{-0.07}$ | 0.14 |
| $714.5^{+1.5}_{-1.5}$ | $3.5^{+1.5}_{-3.5}$* | $11.03^{+0.14}_{-0.23}$ | $713.0^{+0.5}_{-0.5}$ | $4.5^{+0.5}_{-0.5}$ | $11.58^{+0.05}_{-0.03}$ | 3.55 |

* Component is unresolved.
[1] 3 $\sigma$ detection limit.

in our high resolution UES spectra. The component seen at $-7.0$ km s$^{-1}$ in our UES data is split into two components at $-8.9$ km s$^{-1}$ and $-2.5$ km s$^{-1}$ in our higher resolution radio data. This result has been confirmed by observations obtained for the Leiden/Dwingeloo H I survey (Burton 1994).

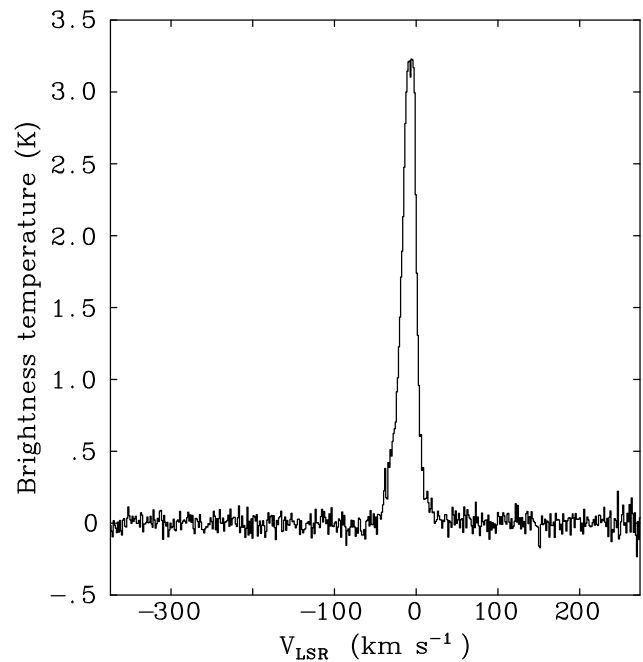

**Fig. 3.** 21 cm profile towards SN 1994D

## 4. Results

We make use of kinematic arguments and the $N(Na^0)/N(Ca^+)$ ratio to determine where in the line of sight to SN 1994D the absorption components are formed. Specifically, calcium is more easily incorporated onto dust grains than sodium; in galactic disks, where gas is generally quiescent, $N(Na^0)/N(Ca^+) \geq 1$, but if the dust grains are disrupted, either by shock waves or cloud collisions, Ca$^+$ is released into the gas phase and $N(Na^0)/N(Ca^+) < 1$. This latter condition is frequently found in clouds moving at high velocity ($\geq 20$ km s$^{-1}$) relative to the local rest velocity, both within a galactic disk and distant from it (Bertin et al. 1993, and references therein). It is also possible to find a low $N(Na^0)/N(Ca^+)$ ratio in gas at low velocity if this gas is warm as Na$^0$ is preferentially removed by collisional ionisation.

**Table 4.** H I gas towards SN1994d

| $v_{lsr}$ (km s$^{-1}$) | FWHM (km s$^{-1}$) | $N(H^0)$ ($10^{18}$ cm$^{-2}$) | $T_{max}$ (K) |
|---|---|---|---|
| $-8.9 \pm 0.1$ | $18.7 \pm 0.2$ | $115.5 \pm 1.7$ | 7540 |
| $-2.5 \pm 0.1$ | $4.7 \pm 0.4$ | $7.3 \pm 0.7$ | 470 |
| $-29.2 \pm 0.9$ | $17.3 \pm 1.7$ | $14.3 \pm 1.6$ | 6520 |

$T_{max}$: upper limit assuming pure thermal broadening of the HI emission

Atomic hydrogen originating in NGC 4526 was generally undetected at 21 cm (Kumar & Thonnard 1983, Giovanardi et al. 1983), with the exception of the observations obtained by Davies & Lewis (1973) which yield a central velocity of +450 km s$^{-1}$, a value frequently quoted in the literature. We do not detect atomic hydrogen near this velocity in our Effelsberg data, or in our low resolution H$\alpha$ or [N II] optical spectra. Our H$\alpha$ spectra yield radial velocities ranging from +625 km s$^{-1}$ at the nucleus of NGC 4526, up to +880 km s$^{-1}$ at the supernova position. A similar value for the central velocity, $v_{\rm hel}$ = +602 km s$^{-1}$ was found by Huchra et al. (1983) and Huchra (1987), from optical spectroscopy as part of the CfA redshift survey ($v_{\rm hel} = v_{\rm lsr}$+ 3.8 km s$^{-1}$). A higher range of radial velocities for the interstellar gas in NGC 4526 is also supported by the CO emission measured by Sage & Wrobel (1989) which peaks at $v_{\rm hel}$ = +743 km s$^{-1}$ and extends from +320 km s$^{-1} \lesssim v_{\rm lsr} \lesssim$ +1060 km s$^{-1}$.

The kinematic evidence suggests that the component present in the high resolution spectra with $v_{\rm lsr}$ = 714 km s$^{-1}$ is produced in NGC 4526; the components at $v_{\rm lsr}$ $\lesssim$ 300 km s$^{-1}$ are however unlikely to originate in that galaxy.

The ratio $N(\text{Na}^0)/N(\text{Ca}^+) \simeq 4$ (Table 3) for this component is a typical value for interstellar gas close to rest velocity in our Galaxy (Siluk & Silk 1974). This, coupled with the good radial velocity agreement suggests that the $v_{\rm lsr}$ = +714 km s$^{-1}$ component originates in rest velocity gas in NGC 4526.

NGC 4526 is a lenticular galaxy classified as S0$_3$ owing to the presence of a nuclear ring of dust (Sandage 1961). Comparison of blue and near-infrared CCD images shows that the dust ring has an angular extent of $\simeq 31'' \times 6''$ (Verón-Cetty & Verón 1988). The detection of molecular gas (Sage & Wrobel 1989) and the fact that NGC 4526 is a bright IRAS galaxy (Soifer et al. 1989) are probably due to the presence of this dust. The line of sight towards SN 1994D, at 7$''$ North and 9$''$ West of the nucleus (Treffers et al. 1994), lies a few arcseconds from the NW border of the ring; the NGC 4526 interstellar gas at +714 km s$^{-1}$ is probably associated with the outskirts of the dust ring.

By adopting the average $N(\text{Na}^0)/N(\text{H})$ ratio of the Milky Way interstellar gas (Ferlet et al. 1985) we estimate $N(\text{H}) \simeq 7.6 \times 10^{19}$ cm$^{-2}$ for the total gas column density in the line of sight within NGC 4526. This value is relatively low for a line of sight which intersects a galaxy close to its nucleus, and is consistent with the low gas content expected for S0 galaxies, which are often ($\simeq$ 80% of the cases) undetected in H I emission (Roberts et al. 1991). Moreover, the absence of multiple components suggests that the structure of the interstellar medium of the host galaxy is relatively simple, at least in the line of sight towards SN 1994D. Clearly care must be taken in extending these results to the galaxy as a whole and, more generally, nents and total column density originating in NGC 4526 may have been much higher if the line of sight had intersected the nuclear ring along its maximum depth rather than at the edge.

### 4.2. High Velocity Clouds

The multiple absorption complex detected in Ca II and Na I at +204 $\lesssim v_{\rm lsr} \lesssim$ +254 km s$^{-1}$ has $N(\text{Na}^0)/N(\text{Ca}^+)$ between 0.1 and 0.4, typical of high velocity gas (Siluk & Silk 1974). The 21 cm counterpart of this gas is not detected in the Effelsberg data at a 3$\sigma$ detection limit $N(\text{H}^0)_{\rm lim} \simeq 7 \times 10^{18}$ cm$^{-2}$. Kumar & Thonnard (1983) however, measured weak emission around $\simeq$ +215 km s$^{-1}$ with $N(\text{H}^0) \simeq 2 \times 10^{18}$ cm$^{-2}$ by pointing the Arecibo 305 m radiotelescope (FHWM = 3.3$'$) in a strip of $\simeq 10' \times 3.3'$ along NGC 4526. The constant radial velocity of this feature along NGC 4526 indicates that the gas is not associated with that galaxy. As discussed previously, the velocities of the multiple absorption complex being $\lesssim$ 300 km s$^{-1}$ indicate it is unlikely they are physically associated with NGC 4526. Furthermore, since this cloud complex has a mean velocity of $\simeq$ $-400$ km s$^{-1}$ with respect to the systemic velocity of NGC 4526 and of $\simeq$ +160 km s$^{-1}$ with respect to that of the Milky Way (v$_{\rm GC}$ = $v_{\rm lsr}$ + 220 sin $l$ cos $b$ km s$^{-1}$), it is more probably related to the surroundings of our Galaxy than to the environment of NGC 4526. The angular dimension of the order of 10$'$ argues against a location in intergalactic space at a distance beyond 1 Mpc, since this would imply that the gas has linear scales of $\approx$ 3 kpc, with a column density of only $2 \times 10^{18}$ cm$^{-2}$, i.e. the gas would have a local density of $\approx 2.2 \times 10^{-4}$ atoms cm$^{-3}$ – even lower than that of the hot phase of the interstellar space where species like Na$^0$ cannot survive. A distance in the range from 1 to 10 kpc, corresponding to a linear size of $\approx$ 3–30 pc, would give a local density of $\approx$ 0.22–0.02 atoms cm$^{-3}$, within the range of normal interstellar cloud sizes and densities.

The cloud complex at +204 $\lesssim v_{\rm lsr} \lesssim$ +254 km s$^{-1}$ cannot be gas corotating in the Milky Way halo, since in this case it would have $|v_{\rm lsr}| \lesssim$ 5 km s$^{-1}$ owing to the relatively small angle between the direction of the supernova ($l$=290$°$) and that of the solar motion around the Galactic centre (see Fig. 4). The complex is not at rest with respect to our Galaxy — if it were at rest it would have $v_{\rm lsr}$ = +70 km s$^{-1}$ as a result of the projection of the solar motion along the line of sight. The gas complex is most likely related to the phenomenon of High Velocity Clouds (HVCs). Although the origin and location of HVCs is not yet fully understood, it is clear that many lie in the surroundings or even in the halo of our Galaxy (Wakker & van Woerden 1991). Only a small number of HVCs were detected by Hulsbosch & Wakker (1988) in a 20$°\times$20$°$ field centered on NGC 4526 at a detection limit of $N(\text{H}^0) \gtrsim 2 \times 10^{18}$ cm$^{-2}$; all were at positive velocities, general span a very wide range of negative and positive velocities, the relatively small velocity difference between the gas at $+204 \lesssim v_{lsr} \lesssim +254$ km s$^{-1}$ and the HVCs in the field suggests that we are dealing with HVCs undetected in the survey of Hulsbosch & Wakker (1988). Our identification of these components as HVCs is supported by the detection of HVCs at remarkably similar velocities (215 – 263 km s$^{-1}$) towards SN 1991T in NGC 4527 (Meyer & Roth 1991), 5° from the SN 1994D sightline – a distance compatible with the size of known HVCs. Alternatively, it is possible that the line of sight to SN 1994D intersects a small Very High Velocity Cloud; several were found by Hulsbosch & Wakker (1988) in the northern sky with velocities in excess of 200 km s$^{-1}$.

Our Ca II and Na I detections, together with the H I detection of Kumar & Thonnard (1983) provide the rare opportunity to compare absorption and emission column densities of HVCs. The ratios $N(Ca^+)/N(H^0)$ and $N(Na^0)/N(H^0)$ of the summed components in the range from +204 to +254 km s$^{-1}$ are extremely high with respect to those found in the ISM of our Galaxy. We find log $N(Ca^+)/N(H^0) = -6.0$ dex, while the highest value from the compilation of interstellar column densities of Stokes (1978) is $-8.1$ dex for disk gas; our value is also higher than that found in more recent work (Albert et al. 1993), concentrating on interstellar absorption towards halo stars. Even considering the lowest Ca$^+$ ionisation fraction found by Phillips et al. (1984), i.e. $N(Ca^{++})/N(Ca^+) \simeq 1$, this implies log $N(Ca)/N(H^0) = -5.7$ dex for the fraction of the total calcium, a value remarkably close to the solar abundance, [Ca/H] = $-5.64$ dex (Anders & Grevesse 1989).

For sodium, we find $N(Na^0)/N(H^0) = -6.8$ dex. In comparison, the average Galactic value is $-8.3$ dex (Stokes 1978) and the solar abundance is $-5.71$ dex.

To bring together these results, we explore three scenarios. Firstly, if we accept the calcium abundance *per se*, then either the gas is of solar metallicity with minimal depletion, or, if there is depletion, the gas must be even more enriched. This proposal is consistent with the below-solar column density of sodium only if the majority of it is present in an unobserved ionisation state such as Na$^+$ – an expected result due to the low ionisation potential of Na$^0$, = 5.14eV. In fact, it is surprising that we see so much Na$^0$ compared to the mean value in the ISM. This suggests we are seeing an ionisation balance for sodium different to that in the local ISM, which may be a symptom of a diluted UV radiation field which is expected far from the Galactic plane.

We must recognise however that the detailed ionisation balance of this gas is largely unknown, and to attribute our results to any single ionisation mechanism is extremely precarious. This leads us to the second possibility: that the gas is, to a significant extent, ionised. In this case, the total hydrogen column density would be substantially, the metallicity of the gas would be lower than the near-solar values suggested in the previous scenario. Enhanced ionisation however fails to explain the excess of neutral sodium compared to that seen in the Galactic ISM.

Lastly, a significant fraction of the hydrogen could be in molecular form, which would again imply that the metallicity of this gas is below solar. However, such an explanation is deemed unlikely since there is no known association between molecular gas and HVCs.

The most likely of these situations appears to be the first: that is, the gas has near solar abundances, with minimal depletion onto dust. This is in agreement with the proposal of Lipman & Pettini (1994), based on observations of Ti II towards halo stars, that gas $\gtrsim$ 1kpc from the Galactic plane contains little dust and has essentially zero depletion.

As mentioned previously, absorption at similiar velocities has been detected towards SN 1991T, a sightline 5° away (Meyer & Roth 1991). These components are an order of magnitude weaker than those we have observed towards SN 1994D; since no absorption was detected by the same authors towards 3C 273, at 1° *further* from the SN 1994D sightline, we conclude that if one HVC complex is responsible for the Ca II absorption in this field, the SN 1991T and 3C 273 sightlines intersect its low column density periphery.

To date there is little information available regarding the abundances found in HVCs. Danly et al. (1993), studied the absorption lines of C II, Si II and O I and report a lower limit for the metallicity of HVC complex M at $Z_{HVC} > -1.0$ to $-1.4$ dex of solar, but predict that the gas is considerably more enriched than this. Bowen & Blades (1993) detected two HVC components towards Mrk 205 using Mg II absorption lines; they found that one had $Z_{HVC} = -2.0$ dex of solar (though this may represent a lower limit due to the possible presence of unresolved high column density components) and for the other they derived a lower limit of $Z_{HVC} = +0.4 \pm 0.5$ dex of solar. Robertson et al. (1991), Songaila (1981) and West et al. (1985) detected Ca II absorption arising from known HVCs; the $N(Ca^+)/N(H^0)$ ratios yield depletions of $-1.2$ to $-2.7$ dex. These values assume an insignificant amount of calcium is present as Ca$^{++}$ and therefore overestimate the depletion. If we assume $N(Ca^{++})/N(Ca^+) \simeq 4$, the mean value in our Galaxy, the depletions are $-0.5$ to $-2.0$ dex.

With such limited information it is impossible to accurately account for the presence of elements in unobserved ionisation states; nevertheless it appears that an appreciable fraction of HVCs may have near-solar abundances, and suffer from little depletion. Observations of different elements and ionisation species will be necessary in order to gain greater insight into the ionisation balance and dust depletion in these clouds more fully, and hence confirm this hypothesis.

with Galactic fountain models (Bregman 1980, Spitzer 1990) in which hot gas is expelled high into the halo and subsequently cools. Such an energetic process would return the refractory elements from the dust grains back to the gas phase, explaining the low depletion of calcium, and could also explain the high overall metallicity and anomalous velocities of the HVC gas. The typical turning height of the fountain models is a few kiloparsecs; as we only see clouds moving away from the Galactic plane and not infalling, this is consistent with our estimation of the distance to the clouds being in the range 1 to 10 kpc.

### 4.3. Milky Way interstellar gas

Fig. 4 shows the expected range of $v_{lsr}$ as a function of distance from the Sun in the direction of SN 1994D assuming circular rotation and the Galactic rotation curve of Brand and Blitz (1991). Clearly, nearly all the components seen in our high resolution spectra have velocities which cannot be explained by the Galactic rotation curve. The two components at $v_{lsr} = -7$ km s$^{-1}$ and $-29$ km s$^{-1}$ are most likely produced in Milky Way or Halo gas. These components are not visible in the spectrum of the foreground stars HD 110411 or HD 110423, setting a lower limit on the distance of 65 pc to absorbing clouds.

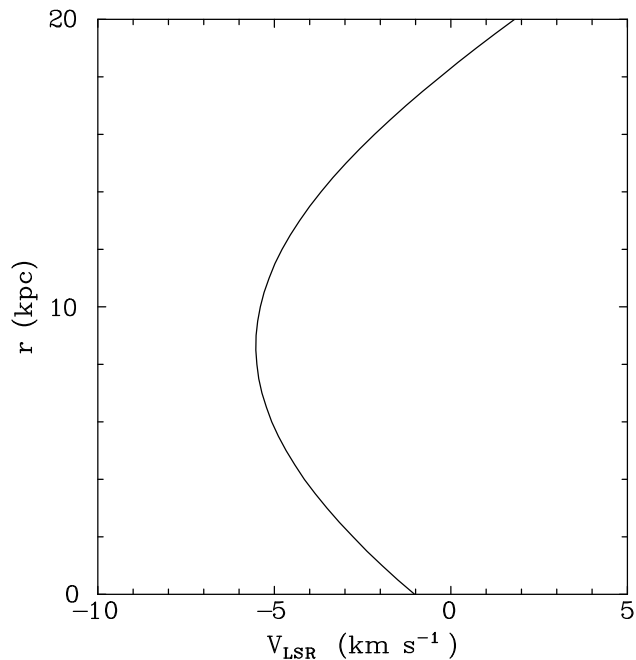

**Fig. 4.** $V_{lsr}$ as a function of distance from the Sun in the direction of SN 1994D assuming the rotation curve of Brand & Blitz (1991) and circular rotation. The effect of rotation parameters deviating from the reference curve can be found in Kaelble et al. (1985)

The component at $-29$ km s$^{-1}$ is detected in Ca II but not in Na I. It is also detected as an asymmetry on the blue wing of the main emission at 21 cm. The width of this 21 cm component indicates that the gas is warm or highly turbulent (Table 4). The low values of the neutral hydrogen and neutral sodium column densities – $N(H^0) \simeq 1.4 \times 10^{19}$ cm$^{-2}$ and $N(Na^0) < 2.5 \times 10^{10}$ cm$^{-2}$ – suggest that we are dealing with warm gas if the correlations between physical parameters and column densities found by Vladilo & Centurión (1994) hold for this component. The ratio $N(Na^0)/N(Ca^+) < 0.19$ is very similar to that found by Bertin et al. (1993) for the warm cloud in the vicinity of the Sun. The ratio $\log N(Ca^+)/N(H^0) = -8.0$ dex is high compared to the typical disk values which are in the range from $-9.7$ to $-8.1$ (see Stokes 1978), but well within the range from $-8.6$ to $-6.5$ found in the Galactic Halo (Albert et al. 1993). Since the interstellar Ca$^{++}$/Ca$^+$ number ratio ranges from 1 to 10 with a mean value of 4 (Phillips et al. 1984), we find $\log N(Ca)/N(H^0) = -7.4 \, (^{+0.4}_{-0.6})$ dex. This implies that calcium is depleted by $-1.8 \, (^{+0.4}_{-0.6})$ dex with respect to the solar abundance value $\log [Ca/H] = -5.64$ dex (Anders & Grevesse 1989). This value for the depletion is relatively low for calcium and is close to the depletion found by Bertin et al. (1993) for the local interstellar cloud. The negative radial velocity of the $-29$ km s$^{-1}$ component cannot be explained in terms of Galactic differential rotation either assuming co-rotation or slower rotation at large distances from the Galactic plane (see Kaelble et al. 1985). The gas at $-29$ km s$^{-1}$ has therefore an intrinsic velocity component directed towards the disk of our Galaxy; it is probably related to the phenomenon of intermediate negative velocities found in the general direction of the North Galactic Pole (Wesselius & Fejes 1973). This gas is likely to be part of the infalling structure located between 250 and 1700 pc above the disk proposed by Danly (1989).

#### 4.3.2. Local gas

The low velocity of the absorption component at $v_{lsr} = -7$ km s$^{-1}$ suggests that it is produced in nearby gas. The ratio $N(Na^0)/N(Ca^+) = 0.5$ however is low for gas close to rest LSR velocity, which has typically $N(Na^0)/N(Ca^+) > 1$ (Siluk & Silk 1974). The low value of the $N(Na^0)/N(Ca^+)$ ratio in this low velocity component can be explained if the gas is warm, as is the case for the local interstellar medium in the solar vicinity (Bertin et al. 1993). The 21 cm emission profiles indicate that the $-7$ km s$^{-1}$ absorption is in fact a blend of two components, one narrow component originating in cold gas and one broader component which may originate in warm gas (see Table 4). The $N(Ca^+)/N(H^0)$ and $N(Na^0)/N(H^0)$ ratios (sum of the two components) are both in the range typical of Galactic clouds, as expected for local gas. The $-7$ km it is not seen in front of the foreground star HD 110411 ($l=295°, b=+73°$) at 65 pc distant, the gas at $-7$ km s$^{-1}$ lies at least 60 pc from the Galactic plane, probably beyond the edge of the Local Bubble (Cox & Reynolds 1987). The gas may be associated with the northern border of Loop I (Berkhuijsen et al. 1971) which is intersected by the supernova line of sight. The tangential distance of Loop I is $70 \pm 40$ pc (Berkhuijsen 1973) consistent with the lower limit for the distance suggested by our data. The line of sight to SN 1994D also intersects the outskirts of the cloud IREC 421 (Désert et al. 1988), which is centered at $l=290.8°$, $b=72.7°$ and has an extension of 3.3 square degrees. Most of the objects in the catalog of Désert et al. (1988) are probably high latitude molecular clouds and the component at $-2.5$ km s$^{-1}$ could represent cold gas associated with IREC 421.

### 4.4. Reddening towards SN 1994D

The $E(B-V)$ reddening of the local gas and of the HVCs can be estimated from the neutral hydrogen column densities by using the the average Galactic ratio $N(H^0)/E(B-V) = 4.8 \times 10^{21}$ cm$^{-2}$ ($\pm 50\%$) of Bohlin et al. (1978). We derive $E(B-V) = 0.029$ and $4.2 \times 10^{-4}$ for the local gas and for the HVCs respectively. As far as the gas in NGC 4526 is concerned we are forced to use the indirect determination of the total hydrogen column density (Sect. 3.1) in conjunction with the relation $N(H_{tot})/E(B-V) = 5.8 \times 10^{21}$ cm$^{-2}$ (Bohlin et al. 1978). This gives $E(B-V) = 0.013$ for the reddening due to the $+714$ km s$^{-1}$ component. The total reddening in front of the supernova is therefore $E(B-V) = 0.046$, with the major source of uncertainty being the NGC 4526 interstellar contribution.

### 5. Summary

The line of sight to SN 1994D in NGC 4526 samples several interstellar media which can be summarised as follows:-

- the interstellar gas in NGC 4526. $N(Na^0)/N(Ca^+)$ is typical of gas at rest in the disk of a galaxy. It has a simple velocity structure, and is most likely associated with the dust ring which surrounds the nucleus of NGC 4526. The estimated total column density is low, as is expected in S0 galaxies. This is the first positive detection of interstellar absorption in a galaxy of early morphological type.
- High Velocity Clouds in the range 1 - 10 kpc from us, with a linear size of $\simeq$ 3-30 pc. The depletion of calcium and sodium in these clouds is less than that normally found in the ISM of our Galaxy, they may be experiencing a diluted radiation field, and they appear to have very low dust content. This is consistent with Galactic fountain models, where refractory elements are returned to the gas phase.
- negative velocities, possibly associated with a structure located between 250 pc and 1700 pc above the Galactic disk.
- Two low velocity components formed in nearby gas. The $N(Na^0)/N(Ca^+)$ ratio of the combined components indicates that the bulk of the gas is warm rather than shocked gas; it is located at least 65 pc from us. One component is formed in cold gas, and may be formed in a high latitude molecular cloud.

*Acknowledgements.* We thank Linda Smith and Max Pettini for obtaining the H $\alpha$ and [N II] observations of NGC 4526.